\documentclass[10pt,aps,prl,twocolumn,superscriptaddress]{revtex4}

\usepackage{epsfig}
\usepackage{dcolumn}
\usepackage{bm}
\usepackage{amsmath}
\usepackage{amsfonts}
\usepackage{latexsym}
\usepackage{amssymb}
\usepackage{color}
\usepackage{hyperref}
\usepackage[normalem]{ulem}

\newcommand{\g}[1]{{\bf #1}}

\newcommand{\eqn}[1]{(\ref{#1})}
\newcommand{\be}{\begin{equation}}
\newcommand{\ee}{\end{equation}}
\newcommand{\bea}{\begin{eqnarray}}
\newcommand{\eea}{\end{eqnarray}}
\newcommand{\ba}{\begin{eqnarray*}}
\newcommand{\ea}{\end{eqnarray*}}
\newcommand{\dagga}{{\phantom{\dagger}}}

\newcommand{\dis}{\displaystyle}
\newcommand{\fract}[2]{\frac{\dis #1}{\dis #2}}

\begin{document} 

\title{Mott physics beyond Brinkman-Rice scenario}

\author{Marcin M. Wysoki\'nski}
\email{mwysoki@sissa.it}

\affiliation{International School for Advanced Studies (SISSA)$,$ via Bonomea 265$,$ IT-34136$,$ Trieste$,$ Italy}
\affiliation{Marian Smoluchowski Institute of Physics$,$ Jagiellonian
University$,$ 
ulica prof. S. \L ojasiewicza 11$,$ PL-30-348 Krak\'ow$,$ Poland}

\author{Michele Fabrizio}
\email{fabrizio@sissa.it}
\affiliation{International School for Advanced Studies (SISSA)$,$ via Bonomea 265$,$ IT-34136$,$ Trieste$,$ Italy}

\date{\today}

\begin{abstract}
The main flaw of the well-known Brinkman--Rice description, obtained through the Gutzwiller approximation,   
of the paramagnetic Mott transition in the Hubbard model is in neglecting high-energy virtual processes 
that generate for instance the antiferromagnetic exchange $J\sim t^2/U$. 
Here we propose a way to capture those processes by combining the Brinkman--Rice approach with a variational Schrieffer-Wolff transformation, and apply this method to study the single-band metal-to-insulator transition in a Bethe lattice with infinite coordination number, where the Gutzwiller approximation becomes exact. We indeed find for the Mott transition a description very close to the real one provided by dynamical mean-field theory; an encouraging result in view of possible applications to more involved models.  
\end{abstract}

\maketitle

A metal to insulator transition driven by the electron-electron repulsion was envisioned by Mott more than fifty years ago~\cite{Mott1961}.
Since then, the underlying physics of this phenomenon has been studied by 
large variety of quantum many-body tools in models for strongly correlated systems~\cite{Brinkman1970,DMFT-Review,Imada1998,Capello2005}.

One of the earliest microscopic descriptions of the 
Mott localisation is owned to Brinkman and Rice~\cite{Brinkman1970}, and obtained 
through the Gutzwiller approximation applied to the half-filled Hubbard model. In their 
scenario the transition to the insulating state occurs when the hopping 
is fully hampered by repulsion, i.e. its expectation value in the variational wavefunction strictly vanishes. 
This result is elegant in many ways. It is fully analytical 
and provides a very intuitive and physically transparent, almost classical, interpretation of the Mott phenomenon.  

Nonetheless, this description, frequently called Brinkman--Rice transition, 
has a severe drawback: the expectation value of the hopping cannot be zero, and it is so in the 
Gutzwiller approximation only because there is a complete, static and dynamic, locking of  
charge degrees of freedom. In reality, dynamical charge fluctuations 
do play a role even deep in the Mott phase, and in particular they mediate the 
antiferromagnetic spin-exchange, as clear by the large $U$ mapping onto the Heisenberg model that 
can be formally derived through the Schrieffer-Wolff transformation~\cite{Spalek1977}. 
 
Since the result of Brinkman and Rice, a variety of quantum many-body tools  have been constructed that are generically able to sensibly capture those dynamical processes~\cite{DMFT-Review,Capello2005,DMRG,Held2007}. In particular, when applied to the Hubbard model, they provide satisfying descriptions of the Mott transition, though relying on heavy numerical computation.
Nowadays, the scenario provided by dynamical mean-field theory (DMFT)~\cite{DMFT-Review}, which becomes exact in infinite dimensions~\cite{Metzner1989}, 
has become an invaluable benchmark to compare with.

In the present work we revisit the problem of the Brinkman--Rice transition, and complement it with the inclusion of the dynamical processes in a semi-analytic manner.
In order to achieve this goal, we construct a method that combines the Gutzwiller's variational approach 
with a variational Schrieffer-Wolff transformation.
As a case study, we apply our technique to the half-filled Hubbard model in the paramagnetic phase on the 
infinitely coordinated Bethe lattice. The energy functional to be minimised can be obtained  fully analytically.   Its minimisation leads to a significantly improved description of the Mott transition as compared to the standard Brinkman--Rice scenario, and much closer to the exact DMFT one~\cite{DMFT-Review}. The improvement is in particular highlighted in: (i) a sizeable 
lowering of the critical interaction strength for a transition; (ii) a lower value of the insulator energy that includes a non-zero expectation value of the hopping $\sim -t^2/U$;  and (iii) 
a proper balance of kinetic and potential energies at the transition. 

The starting point of our analysis is the half-filled single-band 
Hubbard model on the infinitely coordinated Bethe lattice,
\begin{equation}
H= -\frac{t}{\sqrt{z}}\sum_{\langle\g i \g j\rangle}T_{\g i\g j}  
+ \fract{U}{2}\sum_{\g i}\big(n_{\g i\uparrow}+n_{\g i\downarrow}-1\big)^2, \label{HM}
\end{equation}
where ${z\to\infty}$ is the coordination number of the Bethe lattice. 
${T_{\g i\g j}= T_{\g j\g i}
\equiv\sum_\sigma (c_{\g i\sigma}^\dagger c_{\g j\sigma}^\dagga+ 
c_{\g j\sigma}^\dagger c_{\g i\sigma}^\dagga)}$ is the hermitian hopping operator between neighbouring sites $i$ and $j$, and $n_{\g i\sigma}= c_{\g i\sigma}^\dagger c_{\g i\sigma}^\dagga$ the local density of spin ${\sigma=\uparrow,\downarrow}$ electrons.
We rewrite the interaction, last term in Eq.~\eqn{HM}, as   
\begin{equation}
 \begin{split}
H_{\rm int} =\frac{U}{2}\sum_{\g i}\sum_n\,(n-1)^2\;P_{\g i}(n),
 \end{split}
\end{equation}
 where $P_{\g i}(n)$ is the projector at site $\g i$ onto the subspace with $n$ electrons. 
 
 In order to construct a partial Schrieffer-Wolff transformation~\cite{Spalek1977} that accounts for not complete projection of double occupancies, we separately define components of the hopping operator, $T_{\g i\g j}$ projected on the right or on the left onto the configurations where both sites $\g i$ and $\g j$ are singly occupied,
\begin{equation}
\begin{split}
 \tilde  T_{\g i\g j}&\equiv \Big(P_{\g i}(2)P_{\g j}(0)+P_{\g i}(0)P_{\g j}(2)\Big)T_{\g i\g j}\Big(P_{\g i}(1)P_{\g j}(1)\Big),\\
 \tilde  T_{\g i\g j}^\dagger&\equiv\Big(P_{\g i}(1)P_{\g j}(1)\Big)T_{\g i\g j}\Big(P_{\g i}(2)P_{\g j}(0)+P_{\g i}(0)P_{\g j}(2)\Big).
\end{split}
\end{equation}
Their sum is gathered under the form of the new operator $\mathbb{T}_{\g i\g j}\equiv \tilde  T_{\g i\g j}+\tilde  T_{\g i\g j}^\dagger$, while the remaining part of the hopping operator under $\mathcal{T}_{\g i\g j} \equiv T_{\g i\g j}-\mathbb{T}_{\g i\g j}$. We construct the partial Schrieffer-Wolff transformation 
\begin{equation}
\mathcal{U}(\epsilon) = \exp\left(\,\frac{\epsilon}{\sqrt{z}}\;S\right)\,,\label{U-SW}
\end{equation}
through  the anti-hermitian operator 
\begin{equation}
 S=\sum_{\langle \g i\g j\rangle} S_{\g i\g j}=\sum_{\langle \g i\g j\rangle }\big(\tilde  T_{\g i\g j}-\tilde  T_{\g i\g j}^\dagger\big).
\end{equation}
The transformed Hamiltonian reads,
 \begin{equation}
\begin{split}
 \mathcal{H}= &\mathcal{U}(\epsilon)^\dagger H \mathcal{U}(\epsilon) \simeq H- \frac{\epsilon}{\sqrt{z}}[S,H]\\&+\frac{\epsilon^2}{2z}\big[S,[S,H]\big]
-\frac{\epsilon^3}{6z^{3/2}}\Big[S,\big[S,[S,H]\big]\!\Big]\,,
\label{HS}
\end{split}
\end{equation}
where $\epsilon$ is variationally determined so as to minimise the energy.
In the following, we shall assume that for any value of $U$ the optimal $\epsilon$ is small enough to 
safely neglect higher order terms, $\mathcal{O}(\epsilon^4)$ in Eq.~\eqn{HS}. A posteriori, we shall check the validity of such assumption. 
We rewrite the transformed Hamiltonian in a more useful form,
\begin{equation}
\begin{split}
\mathcal{H}&\simeq H+\epsilon\bigg(\frac{t}{z}\sum_{\g i \g j\g k\g l}[S_{\g i \g j}, T_{\g k\g l}]+
\frac{U}{\sqrt{z}}\sum_{\g i \g j} \mathbb{T}_{\g i \g j}\bigg)\\
 &-\frac{\epsilon^2}{2}\bigg(\!\frac{t}{z^{3/2}}\!\sum_{\substack{\g i \g j\g k\\ \g l \g m\g n}}\!\big[S_{\g i \g j},[S_{\g k\g l},T_{\g m\g n}]\big]+\frac{U}{z}\sum_{\g i \g j\g k\g l}[S_{\g i \g j},\mathbb{T}_{\g k\g l}]\!\bigg)\\
&+\frac{\epsilon^3}{6}\bigg(\frac{t}{z^{2}}\!\! \sum_{\substack{\g i \g j\g k\g l\\ \g m\g n\g p\g q}}\! \Big[S_{\g i \g j},\big[S_{\g k\g l},[S_{\g m\g n},T_{\g p\g q}]\big]\!\Big]\\
&\ \ \ \ \ \ \ \ \   \ \ \ \ \ \ \ \ \ \ \ \ \ \ \ \ + \frac{U}{z^{3/2}} \sum_{\substack{\g i \g j\g k\\ \g l\g m\g n}} \big[S_{\g i \g j},[S_{\g k\g l},\mathbb{T}_{\g m\g n}]\big]\bigg)\,,
\label{H}
\end{split}
\end{equation}
 \noindent where we made use of the following equality 
\begin{equation}
 \Big[\sum_{\g i \g j}  S_{\g i \g j},H_{\rm int}\Big]=-U\sum_{\g i \g j}\mathbb{T}_{\g i \g j}.
\end{equation}   

The transformed low energy Hamiltonian \eqref{H} is then analysed by a variational approach.  
Specifically, the ground state of $\mathcal{H}$ is approximated by a variational Gutzwiller wave function $|\psi_G\rangle$ constructed from the uncorrelated Fermi sea $|\psi_0\rangle$ through 
\begin{equation}
\mid\psi_G\rangle\equiv 
\prod_{\g i}\mathcal{P}_{\g i}\mid\psi_0\rangle \label{GA}.
\end{equation}
$\mathcal{P}_{\g i}$ is a linear operator that, in the presence of particle-hole symmetry, can be parametrised as
\begin{equation}
 \mathcal{P}_{\g i}=\sqrt{2}\Big(\sin\theta\ \! P_{\g i}(0)+\cos\theta\ \! P_{\g i}(1)+\sin\theta\ \! P_{\g i}(2)\Big)\,,
\end{equation}
where $\theta$ is a variational parameter bounded by ${\theta\in\{0,\pi/4\}}$, where $\theta=\pi/4$ corresponds to the uncorrelated (metallic) state, whereas $\theta=0$ projects out of $|\psi_0\rangle$ all configurations 
with doubly occupied and empty sites. In other words, the actual variational wavefunction for the ground state of original Hamiltonian $H$  is 
\be
\mid \Psi\rangle = \mathcal{U}(\epsilon)\mid\psi_G\rangle\,,
\label{PSI}
\ee
and depends both on $\theta$ and $\epsilon$. 
The variational energy functional per lattice site (where $N$ is the total number of sites), $\mathcal{F}(\epsilon,\theta)$ can be now obtained as the expectation value in the Gutzwiller wave function \eqref{GA} of the Hamiltonian \eqref{H}, which can be analytically computed in the infinitely coordinated Bethe lattice, 
\begin{equation}
8T_0\,N\,\mathcal{F}(\epsilon,\theta)= \langle\Psi\mid H\mid\Psi\rangle = 
\langle \psi_G\mid\mathcal{H}\mid\psi_G\rangle\,,
\label{f}
\end{equation}
in reduced units of $8T_0$. Here $-T_0$ is the hopping energy per site of $|\psi_0\rangle$, which in a Bethe lattice reads ${T_0 = 8t/3\pi}$.

Already at this point, qualitative differences with respect to the standard Brinkman--Rice transition emerge clearly. In our approach $\theta=0$ providing vanishing double occupancies in $|\psi_G\rangle$ does not yield the same for 
the actual wavefunction $|\Psi\rangle$.  
Explicitly, the density of doubly occupied sites, $d$,  can be calculated as
 \begin{equation}
\begin{split}
 d\equiv& \frac{1}{N}\,\langle \psi_G\mid \mathcal{U}(\epsilon)^\dagger  \,\Big(\sum_{\g i}\,P_{\g i}(2)\Big) \,\mathcal{U}(\epsilon)\mid \psi_G\rangle\,,
\label{d}
\end{split}
\end{equation}
which generically does not provide  $d\!=\!0$ even if ${\theta\!=\!0}$.   

In order to evaluate \eqref{f} as well as \eqref{d} we apply Wick's theorem. 
Each expectation value resulting from this procedure can be 
conveniently visualised by a diagram with nodes denoting sites and 
edges being averages of the inter-site single particle density matrix. 
Sum of diagrams with the same number $x$ of nodes we shall shortly denote as an $x$-vertex. We checked that a satisfying accuracy is obtained by keeping all 
$x$-vertices up to $x=4$. 

The resulting energy functional reads  
\widetext
\begin{equation}
\begin{split}
 \mathcal{F}(\epsilon,\theta)&\simeq\frac{1}{8}\Bigg[  -\Big(1-\frac{\epsilon u}{2\tau}\Big)\sin^2 2\theta  +2u\Big(1-\cos2\theta \Big)
-8\epsilon\, \tau\,\bigg(1-\frac{\epsilon u}{2\tau}\bigg)\, \cos2\theta \\
&-\frac{3\,\epsilon}{8\,\tau}\Bigg(1-\frac{\epsilon u}{4\tau }\Bigg)\sin^2 2\theta\,\cos 2\theta +\epsilon^2\Bigg(\frac{9}{4} -\frac{\epsilon u}{2\tau} \Bigg)\sin^2 2\theta +\frac{32\,\epsilon^3\tau}{3} \cos2\theta  \\
&-\frac{\epsilon^2}{128\,\tau^2}\bigg(1-\frac{\epsilon u}{6 \tau}\bigg)\bigg(-\frac{7}{2}\,\sin^4 2\theta+9\,\sin^2 2\theta\, \cos^2 2\theta\bigg)+ \frac{155 \,\epsilon^3}{192\,\tau} \sin^2 2\theta\, \cos2\theta\Bigg]\,,
\label{fun}
\end{split}
\end{equation} 
\endwidetext
\noindent where interaction strength, $U$ and hopping amplitude, $t$ are rescaled as 
\begin{equation}
 u=\frac{U}{8T_0}, \ \ \ \ \ \ \ \tau=\frac{t}{8T_0}=\frac{3\pi}{64}.
\end{equation}
The first line in Eq. \eqref{fun} includes the 2-vertex contribution, the second line is the 3-vertex one, and finally the third  is the 4-vertex correction. Additionally, the expectation value of the double occupancy $d$ reads 
 \begin{equation}
\begin{split}
 d(\epsilon,\theta&)=\frac{1}{8}\Bigg[ 2 -\bigg(2-4\epsilon^2\bigg)\cos 2\theta\\
&+\frac{\epsilon}{2\tau}\big(1- \epsilon^2 \big)\sin^2 2\theta 
+\frac{3\,\epsilon^2}{32\,\tau^2} \sin^2 2\theta\,\cos 2\theta\\
&+\frac{\epsilon^3}{768\tau^3}\bigg(-\frac{7}{2}\sin^4 2\theta+9\sin^2 2\theta\,\cos^2 2\theta\bigg)\Bigg]\,.
\label{ds}
\end{split}
\end{equation} 
 From \eqref{fun} and \eqref{ds} we can easily recover the results of the standard Gutzwiller approximation applied to the Hubbard model by setting $\epsilon=0$. In this case the Brinkman--Rice transition takes place for $u_{\rm BR}=1$ and the insulating state is characterised by $d=0$.

We search for minima of the functional $\mathcal{F}$ with increasing $u$ by standard methods. Namely, for each $u$ we look for the pair of variables $\{\epsilon,\theta\}$  satisfying 
\begin{equation}
{\partial \mathcal{F}}/{\partial \epsilon}={\partial \mathcal{F}}/{\partial \theta}=0 \,,
\label{condx}
\end{equation}
under the condition that the Hessian is positive definite.
We start observing that for $u=0$, the minimum of the functional \eqn{fun} is correctly determined by $\epsilon=0$ and ${\theta=\pi/4}$ that correspond to fully uncorrelated metal. For interaction strength 
roughly up to $u\simeq 0.4$, the optimised energy is almost coincident with that obtained either by Gutzwiller approximation or by DMFT.

\vspace{-0.9cm}
   \begin{center} 
  \begin{figure}[h!]
     \includegraphics[width=0.48\textwidth]{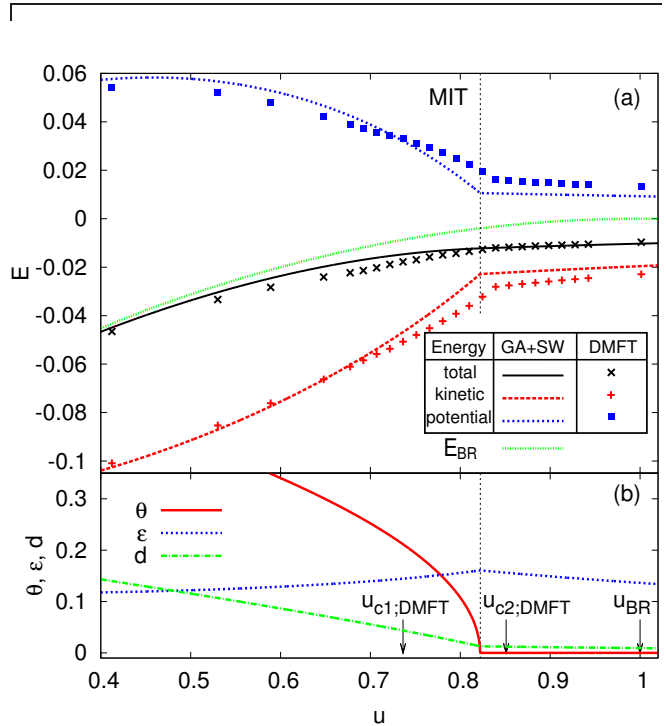}        
\caption{ (a) The equilibrium energy balance across the metal to insulator transition (MIT). For a comparison we have provided data points of the real energies 
from DMFT calculations~\cite{Adriano}. Additionally for a reference we have also included the energy corresponding to the Brinkman-Rice result ($E_{\rm BR}$) \cite{Brinkman1970}.
The transition takes place for quite similar critical interaction as for DMFT ($u_{\rm{c2;DMFT}}$). Also, alike DMFT~\cite{DMFT-Review},  potential and kinetic energies are characterized with the pronounced kinks while the total energy remains smooth.
(b) The equilibrium values of $\theta$, $\epsilon$ and $d$ vs $u$ across metal to insulator transition. We marked the critical values of interaction for a Brinkman--Rice  transition ($u_{\rm BR}$) as well as those obtained by DMFT, in which case at $u_{\rm{c1;DMFT}}$ both, metallic and insulating solutions begin to coexist. Alike DMFT predictions~\cite{DMFT-Review} we obtain non-vanishing double occupancy also in the insulating phase.  }
  \label{fig1}
  \end{figure} 
 \end{center} 

For stronger correlations, $u\gtrsim0.4$, the Gutzwiller approximation starts to deviate 
appreciably with respect to DMFT, while our variational energy remains quite close. 
In Fig.~\ref{fig1}(a) we plot the total energy, as well as separately kinetic and potential energies, of the minimum of functional Eq.~\eqn{fun}, as compared with DMFT \cite{Adriano}, and with the sole Gutzwiller approximation, for which we just show the total energy.

Following Brinkman and Rice \cite{Brinkman1970}, we associate the Mott insulating state with $\theta=0$, which is always a saddle point of the functional \eqn{fun}. However, this saddle point becomes minimum only when metal becomes unstable; the metal to insulator transition is thus continuous and occurs at a critical interaction, $u_c\simeq0.822$, which is sizeably lower than the Brinkman-Rice value, $u_{\rm BR}=1$, and quite close to DMFT, $u_{\rm c2;DMFT}\simeq0.854$. In Fig.~\ref{fig1}(b) we show the values of the variational parameters $\epsilon$ and $\theta$ on the both sides of the transition. In the same figure, we also plot the average double occupancy  $d$ (from Eq.~\eqn{ds}), which is non-zero in the insulating phase and decreases almost linearly in the metallic state. 

In the insulating phase, $u\gtrsim u_c$ when the optimal $\theta=0$, we can analytically calculate 
several quantities. For instance, the saddle point value of $\epsilon$ can be obtained in power series of $\tau/u$: 
\begin{equation}
 \epsilon_{\rm ins}=\frac{\tau}{u}-4\Big(\frac{\tau}{u}\Big)^3+\mathcal{O}\left(\frac{\tau^5}{u^5}\right), \label{eps}
\end{equation}
whereas the energy per site is
\begin{equation}
E_{\rm ins}=-8T_0\Big(\frac{\tau^2}{2u}-\frac{4\tau^4}{3u^3}\Big) +\mathcal{O}\Big(\frac{\tau^6}{u^5}\Big)\simeq-\frac{t^2}{2U}+\frac{4t^4}{3 U^3}.                                                                                  \end{equation}
Additionally, the average double occupancy in powers of $\tau/u$ reads 
\begin{equation}
 d_{\rm ins}=\frac{1}{2}\Big(\frac{\tau}{u}\Big)^2-4\Big(\frac{\tau}{u}\Big)^4+\mathcal{O}\Big(\frac{\tau^6}{u^6}\Big), \label{des}
\end{equation}
which is indeed finite.

 Let us now compare more in detail the above results with the exact DMFT ones ~\cite{DMFT-Review}.  Alike DMFT, we find continuous metal-insulator transition for quite similar critical interaction $U$. However, in our case there is no coexistence region of the insulating and metallic solutions, which in DMFT spreads over significant region ($u_{\rm c1}$ and $u_{\rm c2}$ obtained by DMFT are marked in Fig.~\ref{fig1}(b)). In spite of this deficiency, we do find an energy balance across the transition close to DMFT, and quite different from the Gutzwiller approximation. Indeed,  
hopping and potential energies have kinks at the transition, though the total energy is smooth, and the insulator energy at the leading order scales as $\sim-t^2/2U$ (cf.~Fig.\ref{fig1}(a)). 

In summary, we have analysed a very simple variational wavefunction for a correlated system that consists of a Gutzwiller wavefunction combined with a variational Schrieffer-Wolff transformation. We have benchmarked this wavefunction against the exact DMFT results ~\cite{DMFT-Review} for the paramagnetic Mott transition in the half-filled single-band Hubbard model on a Bethe lattice with infinite coordination number. Although there are obviously differences with exact results, nevertheless our variational wavefunction provides a description of the Mott transition much closer to reality than the Brinkman-Rice scenario. 
More importantly, our wavefunction is able to portray a Mott insulator where charge fluctuations are not completely suppressed as in the Brinkman-Rice scenario, and which therefore has a non-zero expectation value of the hopping. This variational technique might open new possibilities to access Mott physics or related phenomena in more realistic models with minimal computational effort. 

{\it Acknowledgements.}  We are grateful to Adriano Amaricci for
providing us data by DMFT.  
MMW acknowledges support from
Polish Ministry of Science and Higher Education under 
the ``Mobilno\'s\'c Plus'' program, Agreement No. 1265/MOB/IV/2015/0. 
MF acknowledges support from  European Union under the H2020 Framework Programme, ERC Advanced  Grant  No.   692670  ``FIRSTORM".
 

 \end{document}